# Magnetless Circulators Based on Spatiotemporal Modulation of Bandstop Filters in a Delta Topology

Ahmed Kord, *Graduate Student Member, IEEE*, Dimitrios L. Sounas, *Member, IEEE*, and Andrea Alù, *Fellow, IEEE*

*Abstract*— In this paper, we discuss the design rationale and guidelines to build magnet-less circulators based on spatio-temporal modulation of resonant junctions consisting of first-order bandstop filters connected in a delta topology. Without modulation, the junction does not allow transmission between its ports, however, when the natural oscillation frequencies of the constituent *LC* filters are modulated in time with a suitable phase pattern, a synthetic angular-momentum bias can be effectively imparted to the junction and a transmission window opens at one of the output ports, thus realizing a circulator. We develop a rigorous small-signal linear model and find analytical expressions for the harmonic *S*-parameters of the proposed circuit, which significantly facilitate the design process. We validate the theory with simulations and further discuss the large signal response, including power handling and non-linearity, and the noise performance. Finally, we present measured results with unprecedented performance in all metrics for a PCB prototype using a Rogers board and off-the-shelf discrete components.

## I. INTRODUCTION

WIRELESS communications have significantly advanced since the first generation of cellular services was launched in Japan in 1979. Yet, all deployed systems up to date are half-duplex, employing either frequency or time division diplexing for bi-directional communication, therefore limiting the maximum transmission rate allowed by available resources. In a full-duplex system, both the transmitter (TX) and the receiver (RX) operate simultaneously at the same frequency, which, in principle, doubles the capacity of wireless channels [1], [2]. The key challenge in full-duplexing is to have sufficient isolation between the TX and RX nodes (typically >100 dB) to avoid self-interference, i.e., leakage of the strong TX signal into the RX path. Several works from both academic and industrial groups have recently proposed a combination of radio-frequency (RF) and baseband digital signal processing (DSP) techniques [2]-[4] to achieve this goal. RF cancellation is an absolute necessity in full-duplex systems to avoid saturation of the analog-to-digital converters (ADC) and it can be classified into three categories: (i) antenna-based [1]-[3], (ii) circulator-based [3]-[5] and (iii) mixed-signal approaches [6]-[9]. Antenna-based techniques require at least two antennas and are sensitive to their placement, thus weakening the argument for full-duplex as compared to conventional MIMO systems, which can also double the throughput using multiple antennas. Alternatively, mixed signal approaches exploit the fact that the TX signal is already known, and therefore, aim at subtracting it at the RX node. However, what the transceiver actually knows is the clean baseband digital TX signal which becomes very different after it goes through the noisy and non-linear RF chain to be up-converted to the carrier frequency. Therefore, mixed signal approaches, if not carefully designed, may end up adding more interference at the RX node.

The stringent 100 dB specification on self-interference cancellation (SIC) in full-duplex systems can be obtained while using a single antenna by combining DSP and mixed-signal techniques with circulators as depicted in Fig. 1. Nevertheless, circulators come with their own challenge, which is the necessity of breaking reciprocity. In general, this can be achieved using: (i) magnetic-biased anisotropic materials [5], (ii) active devices [10]-[15], (iii) non-linearities [16], [17], or (iv) linear periodically-time-varying (LPTV) circuits [18]-[30]. For decades, non-reciprocity has been almost exclusively achieved through magnetic biasing of ferrite materials, leading to bulky devices, which are incompatible with conventional integrated circuit (IC) technologies. To get rid of magnets, active approaches have been pursued over the years, but they suffer from a fundamentally poor noise figure, limited power handling and small dynamic range, resulting in devices that cannot be deployed in commercial systems. Also nonlinear

Manuscript received on March 31, 2017.

This work was supported by the Qualcomm Innovation Fellowship, the Air Force Office of Scientific Research, the Defense Advanced Research Projects Agency, Silicon Audio, the Simons Foundation, and the National Science Foundation. The authors are with the Department of Electrical and Computer Engineering, University of Texas at Austin, Austin, TX 78712, USA. A. A. is also the Chief Technology Officer of Silicon Audio RF Circulator. (corresponding author: A. A., +1.512.471.5922; fax: +1.512.471.6598; e-mail: alu@mail.utexas.edu).

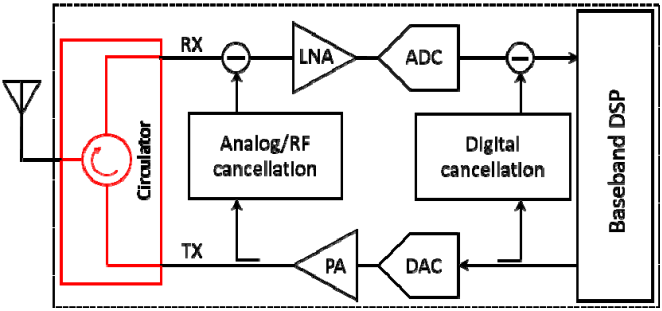

Fig. 1. Circulator in a full-duplex transceiver.

elements have been explored to realize magnet-free non-reciprocal devices, but they inherently lead to signal distortions, and operate only over a limited range of input intensities. Recently, LPTV circuits, based on modulated varactors or banks of switched capacitors [18]-[30], have been presented as a potential alternative towards non-reciprocity, without the drawbacks of the previous approaches.

In particular, [18], [19] presented the idea of parametrically modulating a transmission line (TL) by loading it with varactors and injecting a modulation signal at one port. Such a line allows signal propagation in one direction (the direction opposite to propagation of the modulation signal) as a conventional TL, while in the opposite direction it mixes the injected modulation with the RF signal, upconverting the latter to a different frequency. Such an approach necessitates the TL length to be larger than the wavelength and, more importantly, it requires the use of a diplexer to isolate the two counter-propagating RF signals in the frequency spectrum, thus making it not suitable for integration and less attractive when compared to high performance magnetic-biased circulators. In [24], [25], a fully integrated CMOS magnet-less circulator was presented, where staggered commutation using N-path filters was shown to be equivalent to a highly miniaturized non-reciprocal phase shifter, which, when embedded in a loop of reciprocal phase shifters, can lead to an asymmetric circulator. In this work, the asymmetry of the circuit improved power handling of the TX/ANT path as compared to the ANT/RX path, but it also resulted in sensitivity to impedance mismatches at the RF ports, thus requiring the use of reconfigurable impedance tuners and leading to asymmetric S-parameters. Furthermore, the commutator circuit requires a modulation frequency equal to the RF band's center frequency, which leads to challenges in power consumption in the modulation path, especially in the RF and mm-wave bands, and it also complicates the rejection of any modulation leakage at the RF ports due to switch parasitics.

A related approach for the realization of magnet-less circulators follows the so-called angular-momentum biasing, and it is based on a loop of three resonators modulated in time with 120° phase difference between each other, so that a synthetic angular-momentum is effectively imparted to the circuit [21]-[23]. This approach results in symmetric circulators, which are insensitive to random mismatches, and require a small modulation frequency in the order of 20-30% of the RF band's center frequency, hence reducing power consumption along the modulation path. Furthermore, it allows for easy filtering of the modulation leakage at the RF ports and the opportunity of CMOS implementation via sub-scaled and high-voltage CMOS technologies, which can handle high power. The first circulator based on this approach was presented in [22] by connecting three ladder LC resonators in a delta topology. However, such a connection is non-optimal, as it supports a non-zero common mode that leads to poor matching and an insertion loss in the order of 25 dB, despite the fact that isolation can still be in the order of 40 dB. In [23], it was shown that the common mode problem can be overcome if series LC resonators are connected in a wye topology. One problem with this approach is that it requires a large number of filters in order to prohibit the modulation signal from leaking into the RF ports and vice versa, thus significantly complicating its implementation and increasing insertion loss. Moreover, in this topology, the LC resonators are connected in series with the 50 Ohm port impedance, thus limiting the loaded Q-factor that can be achieved with realistic inductors and in turn the circulator's performance. The circuit reported in [23] showed a measured insertion loss of 9 dB at 130 MHz.

Here, we present a new implementation of angular-momentum circulators that overcomes the mentioned problems of [22], [23]. In particular, we propose connecting parallel LC tanks in a delta topology, which can be considered the dual case of the wye topology in [23]. Interestingly, this circuit does not allow transmission to any port without modulation, however, when angular-momentum is imparted, the degenerate poles of the loop split allowing non-reciprocal transmission with a far more superior performance in all metrics compared to [22], [23]. To the best of our knowledge, this is the first time a bandstop junction is used as the building block to realize circulators, based on any form of biasing, including DC magnets. Like the wye topology in [23], the delta topology presented here does not support a common mode, thus enabling good matching. At the same time, it requires filters and provides a greater control over the loaded Q-factor without the requirement of using impedance transformers at the ports, thus reducing the complexity of the circuit and making it more suitable for integration. In addition, it exhibits better power-handling, since, in contrast to [23], the voltage across the varactors is not amplified by the circuit's resonance (instead, there is an amplification of the current, but this has no effect on the circuit's non-linearity). In this paper, we show and experimentally validate that the proposed circuit results in magnet-less circulators with unprecedented performance, satisfying nearly all metrics of practical systems, e.g., RFID, hence making this work an important step towards the commercialization of full-duplex communications in the near future.

The paper is organized as follows. In Section II, we present the new circuit and develop an analytical model for its linear small-signal response, including the harmonic S-parameters, and describe the design procedure based on this theory. In

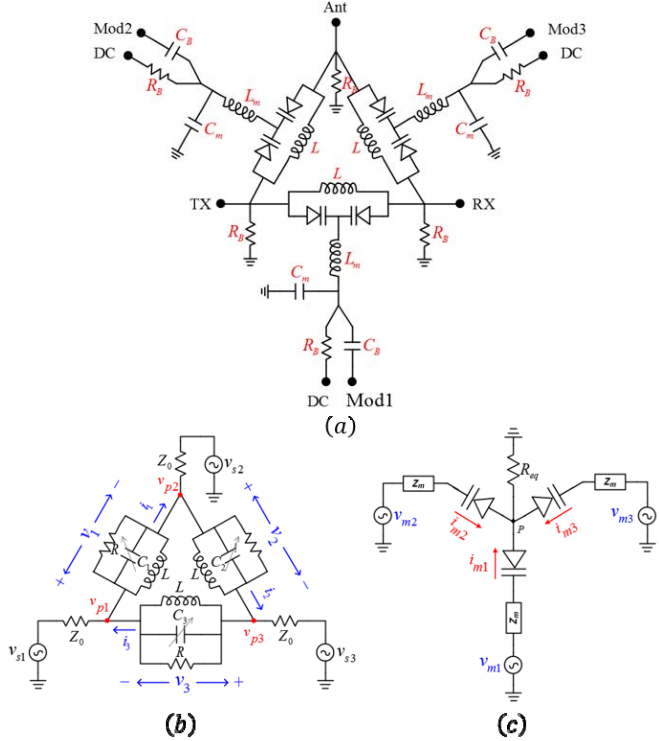

Fig. 2. Proposed magnet-less circulator: (a) Complete schematic. (b) Linear small-signal model at $f_{rf}$. The circuit is connected to sources with a real impedance $Z_0$ and voltages $v_{sn}$, where $n = 1, 2, 3$ is the port index. (c) Simplified model at $f_m$.

Section III, we provide numerical simulations and investigate the circuit's large-signal response, including maximum power handling and non-linearity, followed by a discussion of the noise figure and its associated mechanisms. In Section IV, we provide measured results and discuss the experimental setups for the fabricated prototype in detail. Finally, we draw our conclusions and an outlook on the next steps in this exciting line of research in Section V.

## II. THEORY AND PROPOSED CIRCUIT

### A. Proposed Circuit Topology

Fig. 2 shows the complete implementation of the proposed circuit which consists of three identical parallel $LC$ tanks connected in a loop. These tanks represent first-order bandstop filters with a center frequency given by

$$f_n = \frac{1}{2\pi\sqrt{LC_n}}, \quad n = 1, 2, 3 \tag{1}$$

where $n$ is the tank index, and $L$ and $C_n$ are the total inductance and capacitance of the $n$th tank, respectively. The capacitance $C_n$ is realized via varactors to allow a continuous modulation of the tanks' resonance frequencies, as required in the angular-momentum approach. The modulation signals have the same frequency $f_m$ and amplitude $V_m$, and phase difference $\alpha = 120°$ between different tanks. Varactors are also stacked in pairs to improve power handling and to increase the circuit's 1dB compression point (P1dB). They are also connected in a common-cathode configuration to improve the input-referred third-order intercept point (IIP3) [32]. Modulation is applied to the varactors through matching networks ($L_m$ and $C_m$), which also act as bandstop filters for the RF signal, thus prohibiting its leakage into the modulation ports. An important feature of the proposed circuit is that the modulation signals *see* a virtual ground at the RF ports because of symmetry, therefore alleviating the necessity of using three additional filters as compared to [23] and simplifying the modulation network significantly (Appendix A). Furthermore, the matching networks at the modulation-signal path amplify the modulation voltage, thus relaxing the thus relaxing the requirements on output voltage from the modulation signal generators (Appendix A). Finally, DC biasing is combined with the modulation signals through sufficiently large resistors $R_B$.

### B. Linear Small-Signal RF Analysis

At the RF frequency, the complete circuit in Fig. 2(a) can be simplified as shown in Fig. 2(b), in which the varactors and their DC/modulation network are replaced with time variant capacitors given by

$$C_n = C_0 + \sum_{k=1}^{\infty} a_k v_n^k , \tag{2}$$

where $C_0$ is the static capacitance of the varactors set by the DC bias $V_{DC}$, $a_k$ are the coefficients of a polynomial that models the non-linear $CV$ characteristics of the varactors around their quiescent point, and

$$v_n = v_n^{rf} + v_n^{mod}, \quad n = 1, 2, 3 \tag{3}$$

is the total AC voltage across the $n$th $LC$ tank. $S$-parameters are, by definition, calculated under the small-signal assumption $v_n^{rf} \square v_n^{mod}$, so (2) yields the following expression for the effective capacitance seen by the RF signal:

$$C_n = C_0 + a_1 v_n^{mod} + a_2 (v_n^{mod})^2 + \dots . \tag{4}$$

Assuming weak and linear modulation, so that we can keep terms up to first order, (4) becomes

$$C_n = C_0 + \Delta C \cos(\omega_m t + \varphi_n) , \tag{5}$$

where $\Delta C = aV_m$ is the effective modulation capacitance, $V_m$ and $\omega_m$ are the modulation voltage and frequency, respectively, and $\varphi_n = (n-1)\alpha$, where $\alpha = 120°$ is the constant phase difference between different modulation signals. We also assume that the varactors' and inductor's losses of each tank are combined into a dispersion-less parallel resistance $R$, through which we can define a total "unloaded" quality factor $Q = R / \omega_0 L$. This assumption is acceptable for narrow bandwidths, as is the case in this paper. Equation (5) embeds both the DC bias and the modulation signal into a time-variant capacitance, therefore the tank AC voltages in Fig. 2(b) should be interpreted as the RF signal $v_n = v_n^{rf}$. Also, the RF ports are assumed to have a real impedance $Z_0$, which is typically equal to 50 Ohm.

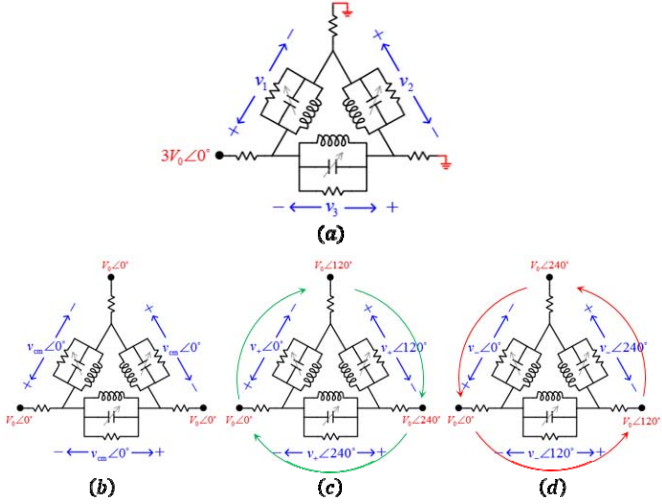

Fig. 3. (a) Excitation at port 1 resulting in the tank voltages $v_n$. (b)-(d) Decomposition of Fig. 5(a) using the superposition of three modes: (b) Common mode, (c) Clock-wise mode, and (d) Counter clock-wise mode.

Applying Kirchhoff's laws to the $n^{\text{th}}$ tank in Fig. 2(b) and writing the result in a matrix form, we get

$$\overline{\overline{A}}\vec{v}'' + \overline{\overline{B}}\vec{v}' + \frac{3Z_0}{L}\vec{v} = \overline{\overline{G}}\vec{v_s}' \ , \tag{6}$$

where $' = \dfrac{d}{dt}$ , $\vec{v} = \{v_1, v_2, v_3\}$ is the tank voltages vector, $\vec{v_s} = \{v_{s1}, v_{s2}, v_{s3}\}$ is the input excitation vector, and $\overline{\overline{A}}$ , $\overline{\overline{B}}$ , and $\overline{\overline{G}}$ are matrices which depend on the circuit elements and modulation parameters as derived in Appendix A. Equation (6) can be written in an even simpler form if, in analogy with magnetic circulators, we express $v_n$ as a superposition of three modes, i.e., $v_n = v_{\text{cm}} + v_+ e^{j(n-1)\alpha} + v_- e^{-j(n-1)\alpha}$ , where $v_{\text{cm}}$, $v_+ e^{j(n-1)\alpha}$, and $v_- e^{-j(n-1)\alpha}$ are defined as the common, clockwise and counter-clockwise modes, respectively. The common mode here refers to the in-phase signal component of the voltages $v_n$. The clockwise mode refers to the signal component whose phase increases in a clockwise direction by $\alpha = 120°$, similar to a TL wave propagating in that direction. Similarly, the counter-clockwise mode refers to a propagating wave in the opposite direction, i.e., a signal whose phase increases counter-clockwise. As an example, an excitation at port 1 as shown in Fig. 3(a), where $V_0$ is a constant amplitude, can be decomposed using superposition into these predefined modes as shown in Fig. 3(b)-(d), while noticing that $1 + e^{j2\pi/3} + e^{j4\pi/3} = 0$ . More generally, for excitation at all ports, this can be expressed through the matrix transformation

$$\vec{v} = \overline{\overline{T}}^{-1}\overline{v} \ , \tag{7}$$

where $\vec{v} = \{v_{\text{cm}}, v_+, v_-\}$ is the vector of the defined mode voltages and the operator $\overline{\overline{T}}$ is given by

$$\overline{\overline{T}} = \begin{bmatrix} 1 & 1 & 1 \\ 1 & e^{j\alpha} & e^{-j\alpha} \\ 1 & e^{j2\alpha} & e^{-j2\alpha} \end{bmatrix} \ . \tag{8}$$

Applying this transformation and considering the case when only port 1 is excited, as in Fig. 3, i.e., $\overline{v_s} = \{1,0,0\}$ , we get

$$3Z_0 C_0 v_{\text{cm}}'' + v_{\text{cm}}' + \frac{3Z_0}{L} v_{\text{cm}} = 0, \tag{9}$$

$$3Z_0 C_0 v_+'' + \frac{3}{2} Z_0 \Delta C e^{-j\omega_m t} v_-'' + \left(1 + \frac{3Z_0}{R}\right) v_+' \\ - \frac{j3}{2} Z_0 \Delta C \omega_m e^{-j\omega_m t} v_-' + \frac{3Z_0}{L} v_+ = \frac{1}{6}(3 - j\sqrt{3}) v_{s1}', \tag{10}$$

$$3Z_0 C_0 v_-'' + \frac{3}{2} Z_0 \Delta C e^{j\omega_m t} v_+'' + \left(1 + \frac{3Z_0}{R}\right) v_-' \\ + \frac{j3}{2} Z_0 \Delta C \omega_m e^{j\omega_m t} v_+' + \frac{3Z_0}{L} v_- = \frac{1}{6}(3 + j\sqrt{3}) v_{s1}'. \tag{11}$$

Notice that a general $\overline{v_s}$ can be constructed using a linear superposition of individual port excitations and that excitation from the other ports can be inferred from excitation at port 1 based on the circuit's symmetry. Equation (9) has only the trivial solution $v_{\text{cm}} = 0$ . This means that the common mode is not excited and that the input power is split between the counter-propagating modes $v_+$ and $v_-$ , which are given by the coupled differential equations (10) and (11). These equations can be solved by applying the Fourier transform, yielding (Appendix B)

$$V_{\pm}(\omega) = \sum_{k=-1}^{1} H_k^{\pm}(\omega) V_{s1}(\omega + k\omega_m), \tag{12}$$

where

$$H_0^{\pm}(\omega) = \frac{\frac{j\omega}{6}(3 \mp j\sqrt{3})}{D_{\pm}(\omega)}\left( \begin{array}{l} -3C_0 Z_0 (\omega \pm \omega_m)^2 + \frac{3Z_0}{L} \\ + j(\omega + \omega_m)\left(1 + \frac{3Z_0}{R}\right) \end{array} \right), \tag{13}$$

$$H_{-1}^-(\omega) = \frac{\frac{j}{4}(3 - j\sqrt{3})\Delta C Z_0 \omega(\omega - \omega_m)^2}{D_-(\omega)}, \tag{14}$$

$$H_1^+(\omega) = \frac{\frac{j}{4}(3 + j\sqrt{3})\Delta C Z_0 \omega(\omega + \omega_m)^2}{D_+(\omega)}, \tag{15}$$

$$H_{-1}^+(\omega) = H_1^-(\omega) = 0. \tag{16}$$

The denominator function $D_{\pm}(\omega)$ is given by

$$V_1(\omega_k) = V_+(\omega_k) + V_-(\omega_k), \tag{21}$$

$$V_2(\omega_k) = e^{j2\pi/3}V_+(\omega_k) + e^{-j2\pi/3}V_-(\omega_k), \tag{22}$$

$$V_3(\omega_k) = e^{-j2\pi/3}V_+(\omega_k) + e^{j2\pi/3}V_-(\omega_k), \tag{23}$$

where $\omega_k = \omega + k\,\omega_m$ and $k = -1,0,1$ is the harmonic index. Using Kirchhoff's laws, the tank currents $I_n(\omega_k)$ can be calculated from the voltages $V_n(\omega_k)$ (see Appendix B), then the harmonic $S$-parameters for excitation at port 1 can be found as follows:

$$
\begin{aligned}
S_{11}(\omega_k,\omega) &= \delta_{kl} - 2\frac{Z_0\left[I_1(\omega_k) - I_3(\omega_k)\right]}{V_{s1}(\omega)}, \\
S_{21}(\omega_k,\omega) &= 2\frac{Z_0\left[I_2(\omega_k) - I_3(\omega_k)\right]}{V_{s1}(\omega)}, \\
S_{31}(\omega_k,\omega) &= 2\frac{Z_0\left[I_1(\omega_k) - I_2(\omega_k)\right]}{V_{s1}(\omega)}.
\end{aligned}
\tag{24}
$$

If $k = 0$, $\omega_k = \omega$ and (24) give the conventional $S$-parameters defined for linear time-invariant (LTI) systems which relate the input and output powers at the same frequency, while $k \neq 0$ gives the transformation of the input signal at $\omega$ to the intermodulation products at $\omega_k = \omega \pm \omega_m$. In the next sections, the $S$-parameters refer to the conventional ones, unless stated otherwise. Also, due to the circulator's threefold rotational symmetry, the rest of the $S$-parameters can be found by rotating the indices as $(1,2,3) \to (2,3,1) \to (3,1,2)$, resulting in

$$
\begin{aligned}
S_{11} &= S_{22} = S_{33} \\
S_{21} &= S_{32} = S_{13} \\
S_{31} &= S_{12} = S_{23}
\end{aligned},
\tag{25}
$$

Similarly, the harmonic input impedance at port 1 is found by

$$Z_{in}(\omega_k,\omega) = \frac{V_{p1}(\omega)}{I_{p1}(\omega_k)} = \frac{V_{s1}(\omega)}{I_1(\omega_k) - I_3(\omega_k)} - Z_0. \tag{26}$$

Like for the $S$-parameters, $k = 0$ yields the conventional input impedance, which relates voltages and currents having the same frequency. Generally, if $k \neq 0$, $Z_{in}(\omega_k,\omega)$ corresponds to a trans-impedance which relates the input voltage at $\omega$ to the excited current at $\omega_k$. It is also worth mentioning that

$$Z_{in}(\omega_k,\omega) = Z_0 \frac{1 + S_{11}(\omega_k,\omega)}{1 - S_{11}(\omega_k,\omega)}$$ still holds for all $k$.

Equations (24), (25) give the $S$-parameters as a function of the circuit elements $L$, $C_0$, $Q$ (recall that $Q$ is the unloaded quality factor of the $LC$ resonators, incorporating both inductor's and varactor's losses) and the modulation parameters $f_m$ and $\Delta C$. These parameters should be chosen for operation around a given frequency $f_{rf}$ while achieving certain specifications on insertion loss (IL), return loss (RL),

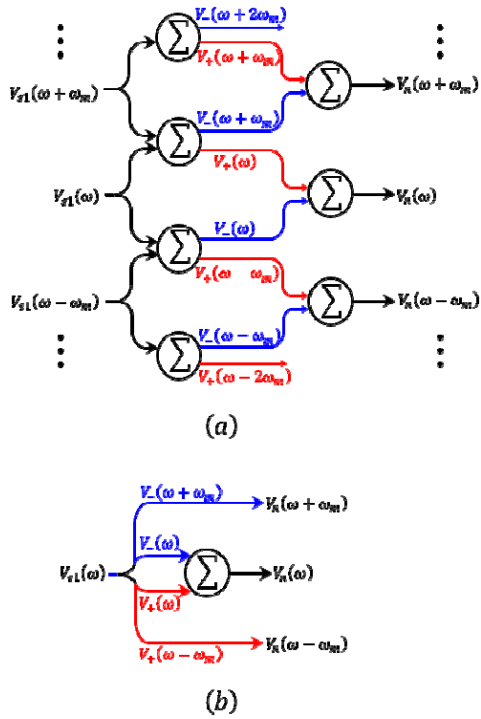

$(a)$

$(b)$

Fig. 4. (a) Frequency mixing map for wideband signals. (b) Frequency mixing map for narrowband signals.

$$
\begin{aligned}
D_\pm(\omega) = &-\frac{9}{4}\Delta C^2 Z_0^2 \omega^2 \left(\omega \pm \omega_m\right)^2 \\
&+ \left(-3C_0 Z_0 \omega^2 + \frac{3Z_0}{L} + j\,\omega\left(1 + \frac{3Z_0}{R}\right)\right) \\
&\times \left(-3C_0 Z_0\left(\omega \pm \omega_m\right)^2 + \frac{3Z_0}{L} + j\left(\omega \pm \omega_m\right)\left(1 + \frac{3Z_0}{R}\right)\right).
\end{aligned}
\tag{17}
$$

Equation (12) shows that the voltage at any frequency $\omega$ depends on the input voltages at different frequencies, in particular the frequencies $\omega$, $\omega + \omega_m$ and $\omega - \omega_m$, as a result of the modulation. A graphical representation of the frequency mixing mechanism is provided in Fig. 4(a). If the input signal is monochromatic with frequency $\omega$, (12) and Fig. 4(a) show that the voltages have three components at frequencies $\omega$, $\omega + \omega_m$ and $\omega - \omega_m$, as shown graphically in Fig. 4(b), with amplitudes

$$\frac{V_\pm(\omega)}{V_{s1}(\omega)} = \frac{j\omega}{6}\left(3 \mp j\sqrt{3}\right)\frac{\left(-3C_0 Z_0\left(\omega \pm \omega_m\right)^2 + \frac{3Z_0}{L}\right.}{D_\pm(\omega)}\left.+ j\left(\omega \pm \omega_m\right)\left(1 + \frac{3Z_0}{R}\right)\right), \tag{18}$$

$$\frac{V_\pm(\omega \mp \omega_m)}{V_{s1}(\omega)} = \frac{\frac{j}{4}\left(3 \pm j\sqrt{3}\right)\Delta C Z_0\left(\omega \mp \omega_m\right)\omega^2}{D_\mp(\omega)}, \tag{19}$$

$$V_\pm(\omega \pm \omega_m) = 0. \tag{20}$$

Next, in order to find the $S$-parameters, (18)-(20) are transformed back to the tank voltages $\bar{V} = \{V_1,V_2,V_3\}$. In particular, using the inverse transformation of (7) and recognizing that $V_{cm} = 0$, we find isolation (IX) and BW. Fig. 5 shows the *S*-parameters and the fractional bandwidth versus the normalized modulation frequency $f_m/f_{rf}$ and normalized modulation amplitude $\Delta C/C_0$ for $f_{rf} = 1$ GHz, $C_0 = 7.67$ pF, $L = 3.4$ nH and $Q = 70$, which is a reasonable value for a PCB design. The values of $L$ and $C_0$ were selected by assuming an initial value for one of them and finding the other such that the resonance frequency $f_0 = 1/\left(2\pi\sqrt{LC_0}\right)$ occurs at the design frequency $f_{rf}$. The circulator's bandwidth is defined as

$$\text{BW} = \min\left\{\Delta f\big|_{\text{IL}>\beta}, \Delta f\big|_{\text{IX}>\gamma}\right\}, \qquad (27)$$

where $\Delta f\big|_{\text{IL}>\beta}$ and $\Delta f\big|_{\text{IX}>\gamma}$ are the frequency ranges where insertion loss (IL) is less than $\beta$ dB and isolation (IX) is more than $\gamma$ dB, respectively. For the results in Fig. 5(d), we choose $\beta = 4$ dB and $\gamma = 20$ dB, in which case, one can find that $\Delta f\big|_{\text{IL}>4\text{dB}}$ is always greater than $\Delta f\big|_{\text{IX}>20\text{dB}}$, i.e., the bandwidth is determined by the minimum isolation specification. Fig. 5 allows us to find the modulation parameters for which any of the *S*-Parameters or the bandwidth become optimum (minimum $S_{11}$, maximum $S_{21}$, minimum $S_{31}$ and maximum BW). These quantities do not necessarily become optimum under the same modulation parameters, showing that we may need to trade off one or several of them when we select the modulation parameters. In this paper, we give priority to isolation and we choose the circuit to operate at point $p_3$, where isolation at the design frequency is maximum. The various circuit parameters for this operation point are summarized in Table I. Other choices that give priority to other metrics, or a combination of them, depending on the design specifications, are equally valid. For example, if maximizing isolation at the center frequency is not important, one could select point $p_4$ as the operation point, where the 20 dB isolation bandwidth is maximum. Notice that point $p_4$ is also close to point $p_2$, where insertion loss is minimum. In general, if the generated charts do not provide a solution that meets the given specifications, one can try different values of $L$ and $C_0$ that satisfy $f_0 = f_{rf}$ until reaching the desired solution.

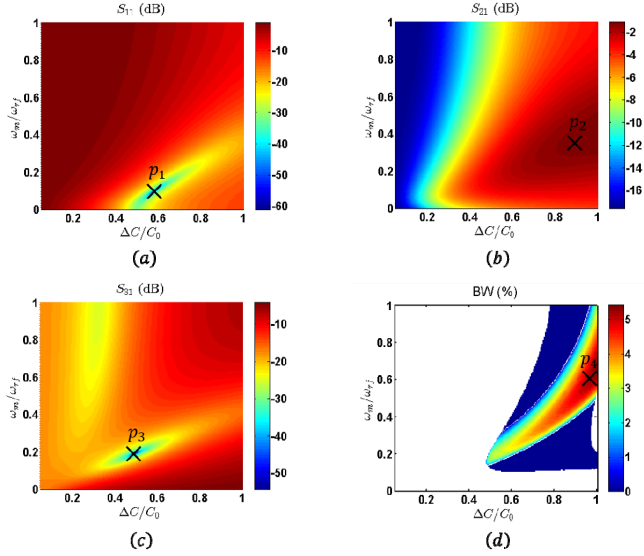

Fig. 5. *S*-parameters at $f = 1$ GHz versus modulation parameters for $f_{rf} = 1$ GHz, $C_0 = 7.67$ pF, $L = 3.4$ nH and $Q = 70$. (a) Return loss. (b) Insertion loss. (c) Isolation. (d) Fractional bandwidth.

TABLE I
THEORETICAL DESIGN PARAMETERS

| Element | Value |
|---------|-------|
| $Q$ | 70 |
| $L$ | 3.4 nH |
| $C_0$ | 7.67 pF |
| $\Delta C / C_0$ | 0.46 |
| $f_m$ | 190 MHz |

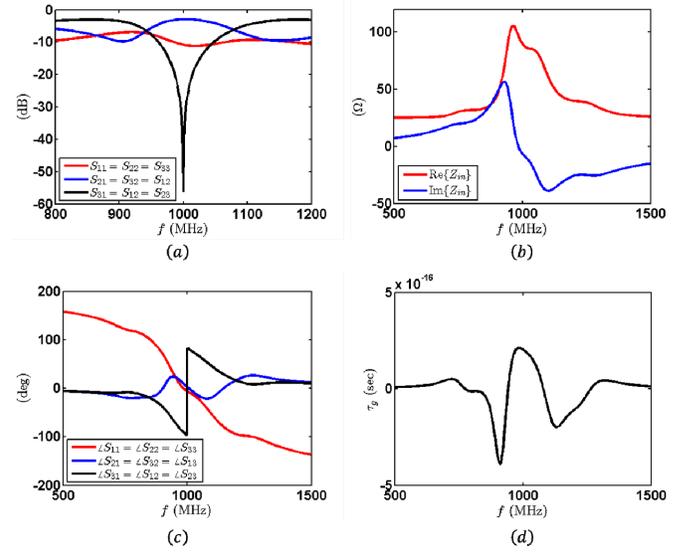

Fig. 6. Theoretical results: (a) *S*-parameters magnitude. (b) *S*-parameters phase. (c) Input impedance. (d) Transmission group delay.

Fig. 6(a) shows the *S*-parameters in dB using (18)-(25) and the values given in Table I. Insertion loss, return loss and isolation at the center frequency of 1 GHz are 2.9 dB, –10.8 dB and 56 dB, respectively, and the fractional BW based on the definition of (27) is 2.7% (27 MHz). Input impedance is also shown in Fig. 6(b) which is almost real, but not equal to 50 Ohm (the port impedance), at the design frequency $f_{rf}$. This is related to the fact that the circuit was not chosen to operate at point $p_1$ in Fig. 5(a), where the return loss becomes very small. Impedance matching can be improved by using simple *LC* matching networks at the RF ports at the expense of increasing the form factor and the overall resistive loss [31]. Fig. 6(c) and Fig. 6(d) also show the phase response and the transmission group delay $\tau_g = -\dfrac{d\angle S_{21}}{d\omega}$, respectively. Clearly, $\tau_g$ is almost flat in the 2.7% operational band, thus imposing minimal dispersion to the transmitted signal through the circulator and allowing its use with any coherent or non-coherent digital modulation scheme, as in practical communication systems.

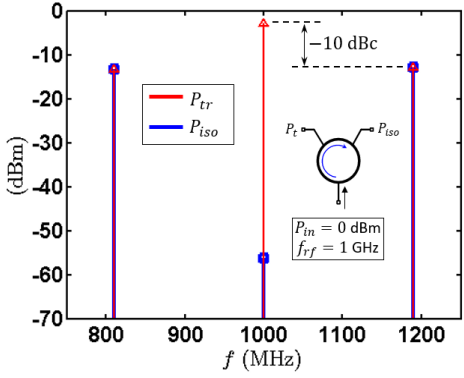

Fig. 7. Simulated harmonic spectrum at the transmitted and isolated ports for a single tone input at $f_{rf} = 1$ GHz and $P_{in} = 0$ dBm.

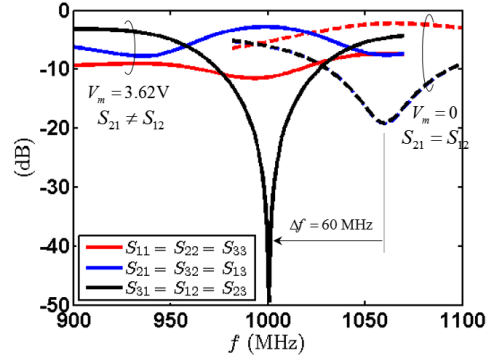

Fig. 8. Simulated $S$-parameters: without modulation (dashed lines) and with modulation (solid lines).

TABLE II
List of the Discrete Components Used in Our Design

| Element | Value |
|---------|-------|
| $D$ | ~8 pF @ VDC=21.6V |
| $L$ | 3.3 nH |
| $L_m$ | 72 nH |
| $C_m$ | 24 pF |
| $R_B$ | 100 KOhm |
| $C_B$ | 1000 pF |

Fig. 7 shows the circulator's harmonic response at the transmitted and isolated ports, i.e., $P_t$ and $P_{iso}$, respectively, for a monochromatic input at $f_{rf} = 1$ GHz and $P_{in} = 0$ dBm. Notice that the presented circulator is rotationally symmetric as mentioned earlier, hence Fig. 7 results apply for excitation at any of the RF ports where, for example, if the input is incident from port 1, then $P_t$ and $P_{iso}$ become $P_2$ and $P_3$, respectively. The IM products are only –10 dBc, where dBc is a normalized unit with respect to the carrier (center) frequency, which poses an interference problem to neighboring channels and may saturate the RX front-end, especially at high TX power, even if the fundamental harmonic is sufficiently attenuated. It can be shown either in simulations or from the analysis in Section II.B that these products are reduced when using a larger modulation frequency, yet at the expense of increasing the power consumption in the modulation path. Further reduction is possible using channel or band pre-selection filters though, again, this would put a restriction on the minimum modulation frequency, hence on the overall power consumption, to relax the requirements on the sharpness of these filters.

## III. Simulation Results

Based on the theoretical results presented in the previous section, a PCB circulator operating at 1 GHz was designed using off-the-shelf discrete components, as listed in Table II. These components were chosen based on the design parameters in Table I. The layout on a Rogers board was simulated using ADS Momentum and the generated $S$-parameters were combined with the rest of the circuit components in ADS to perform post-layout circuit/EM co-simulations. By following such approach, we take into account all parasitics, either due to the finite length of the interconnecting transmission lines or the pads of the components' footprints. Commercially available measured $S$-parameters of the passive $RLC$ elements and the full non-linear spice model with package parasitics of the varactors were also used.

### A. S-parameters

Fig. 8 shows the simulated $S$-parameters with and without modulation for $V_{DC} = 21.6$ V. Without modulation ($V_m = 0$), the circuit is, clearly, reciprocal with the same transfer function seen between the input port, e.g. port 1, and any of the output ports, i.e., $S_{21} = S_{31}$. As mentioned before, the proposed circulator without modulation is a bandstop filter, i.e., at the resonance frequency transmission is not allowed towards any of the output ports, with the input power mostly reflected back and residual transmission is merely due to the finite quality factor of the constituent $LC$ tanks. Fig. 8 shows that indeed without modulation, transmission is –20 dB at the unmodulated center frequency of 1.06 GHz. When spatio-temporal modulation is applied with $f_m = 190$ MHz and $V_m = 3.62$ V, the $S$-parameters become non-reciprocal: for excitation from port 1, 2 and 3 the power is mostly transmitted to ports 2, 3, 1, respectively. We also notice that the center frequency is shifted down to 1 GHz, because of the varactors' second-order non-linearity, as will be explained later. The achieved insertion loss, return loss and isolation at 1 GHz are 2.8 dB, –11.34 dB and 55 dB, respectively, which are all in agreement with the linear small-signal analytical results of Section II. On the other hand, the fractional BW decreases to 1.8% (18 MHz), because of all the parasitics which were neglected in the theoretical analysis. The simulated phase response and group delay are also shown in Fig. 9.

Fig. 10(a) presents the $S$-parameters for different DC voltages, showing that the circulator can be reconfigured to operate at different frequencies spanning a range of 100 MHz, which is about six times larger than the instantaneous bandwidth, by simply controlling the DC bias and adjusting the modulation voltage to maintain the same $S$-parameters. This range can be extended further if the modulation frequency is also controlled, e.g. using voltage controlled oscillators (VCO) to generate the modulation signals.

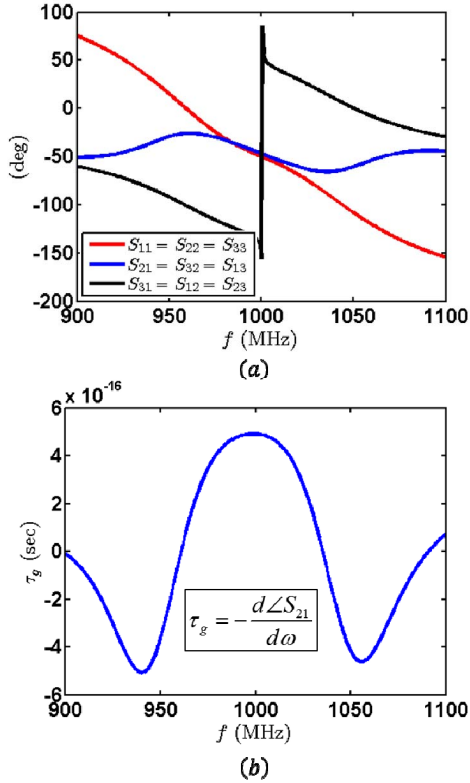

Fig. 9. (a) Simulated phase response. (b) Simulated group delay.

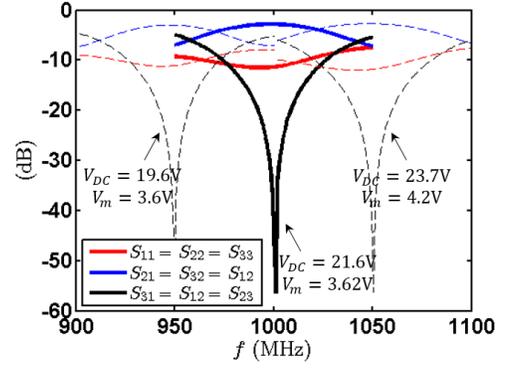

Fig. 10. Simulated reconfigurable $S$-parameters.

### B. Power Handling and Non-Linearities

Non-linearity in the circuit is predominantly due to the use of varactors, related to the non-linear $CV$ characteristics as given by (2) and the finite forward-conduction and breakdown voltages, $V_f$ and $V_B$, respectively. In order to maximize the power handling of a given varactor, the biasing should be such that the circulator operates in the middle channel of the tunability band at $V_{DC} = \frac{V_f + V_B}{2}$, which is approximately equal to 20 V for the varactors used in this paper. Clearly, as the DC bias deviates from this optimal value, e.g. as we tune the circulator for operation at a different channel, the maximum power that the varactor can handle decreases. This requires to set a back-off limit from both $V_f$ and $V_B$ in order to maintain, approximately, the same linearity performance in all channels.

In order to investigate the second factor that contributes to the non-linear characteristics of the circuit, i.e., the non-linear $CV$ curve of the varactors around the quiescent point, we invoke (2) and substitute for $v_n = V_m \cos(\omega_m t) + V_{rf} \cos(\omega_{rf} t)$, which yields

$$C = b_0 + \sum_{k=1}^{\infty} \left[ \begin{array}{l} b_k \cos(k\,\omega_m t) + d_k \cos(k\,\omega_{rf} t) \\ + \sum_{l=-\infty}^{\infty} e_{lk} \cos\left[\left(k\,\omega_m + l\,\omega_{rf}\right)t\right] \end{array} \right], \qquad (28)$$

where $b_k$, $d_k$ and $e_{lk}$ are polynomial coefficients, which up to the first-order term are given by

$$\begin{aligned} b_0 &= C_0 + \frac{1}{2}a_2\left(V_m^{\,2} + V_{rf}^{\,2}\right) + \cdots, \\ b_1 &= aV_m + \frac{3}{4}aV_m^{\,3} + \cdots, \\ d_1 &= aV_{rf} + \frac{3}{4}aV_{rf}^{\,3} + \cdots. \end{aligned} \qquad (29)$$

Notice that for simplicity, the subscript $n$ was dropped and the sinusoidal phases were assumed zero since they are irrelevant to this analysis. The DC term $b_0$ represents the effective static capacitance in the presence of large modulation and RF signals while taking into account the varactors' non-linearities. We notice a shift in such capacitance with respect to the static value $C_0$ due to the second order non-linearity. Generally, all even order terms in (2) would result in a similar shift but the second order one is predominant. For a convex $CV$ curve, as for the majority of commercial varactors, $a_2$ is positive, therefore the resonance frequency is shifted "down" in agreement with the result in Fig. 8. For $V_{rf} \Box V_m$, this shift is constant and independent of the RF signal, however, when $V_{rf}$ becomes sufficiently large, $b_0$ changes with the input power, or in other words, the varactor is compressed. Clearly, when this occurs, both IL and IX at the fixed design frequency, e.g. 1 GHz, will also compress.

The rest of the terms in (28) correspond to harmonic variation of the varactor's capacitance at different frequencies. For example, $b_1$ corresponds to the effective capacitance variation at the imposed modulation frequency $\omega_m$ which, in contrast to (5), is a non-linear function of the modulation amplitude $V_m$ due to the non-linear characteristics of the varactors' $CV$ curve. The term $d_1$ corresponds to a harmonic capacitance variation at the RF frequency $\omega_{rf}$, effectively corresponding to modulation of the circuit with a modulation frequency $\omega_{rf}$. Since such a modulation frequency is very far from the optimum modulation frequency $\omega_m$, this term has a negligible effect on the circuit's response. Similarly, the other terms in (28) correspond to modulation of the circuit at other individual or mixed harmonics of $\omega_m$ or $\omega_f$ and for the same reason as for $d_1$, they have a negligible effect on the circuit's response.

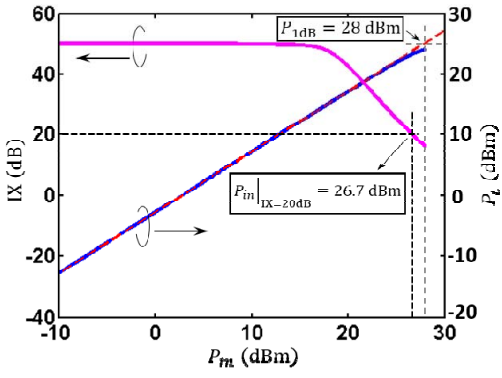

Fig. 11. Simulated compression point and maximum power handling.

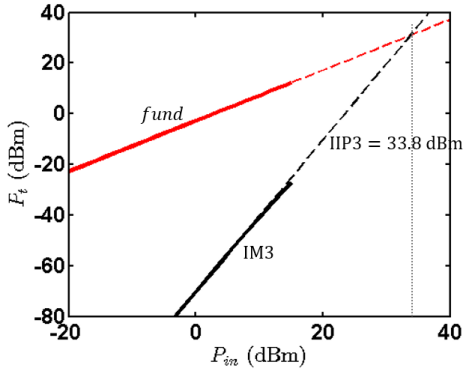

Fig. 12. Simulated two-tone test and achieved IIP3.

Fig. 11 shows the output power at the transmitted port and the isolation versus the input power, assuming that the input frequency is equal to the design frequency of 1 GHz. We can see that both the output power and the isolation compress after a certain input power, due to the non-linear characteristics of the varactors, as explained before. The maximum allowable input power level is defined as

$$P_{\max} = \min\{P_{in}|_X, P_{in}|_{IX>\sigma}\}, \quad (30)$$

where $P_{in}|_X$ is the input $X$ dB compression point and $P_{in}|_{IX>\sigma}$ is the maximum input power to maintain $\sigma$ dB isolation (IX) at the center frequency. For the typical values: $X = 1$ dB and $\sigma = 20$ dB, Fig. 11 shows that the achieved $P_{\max}$ is 26.7 dBm. Such a large value exceeds by several orders of magnitude what conventional active approaches can achieve either in PCB or IC platforms [10]-[14]. It is also worth mentioning that this large compression point is partially due to the stacking of the varactors in pairs, as in Fig. 2(a). Such an approach halves the RF voltage on each varactor, thus increasing the maximum allowable input power by 3 dB, approximately. In general, stacking $N$ varactors can improve the compression point by $\sim 20\log_{10}(N)$, but it also complicates the modulation network and, more importantly, increases the insertion loss, due to the parasitic loss of the varactors. Finally, the circuit is also tested with two in-band input tones at $1000 \pm 0.5$ MHz, as shown in Fig. 12 where IIP3 is found to be 33.8 dBm. Notice that the difference between IIP3 and P1dB is not necessarily 9.6 dB as expected in third-order time-invariant systems [33], due to the time-varying characteristics of the circuit.

### C.  Noise Figure

Noise performance of the ANT/RX path is another critical metric for the performance of the circulator, since it comes at the forefront of the receiving path superseding the low noise amplifier (LNA); therefore, it is highly desirable that the circulator adds minimal noise in order not to limit the overall signal-to-noise ratio (SNR) of the transceiver. In the following discussion, we detail the different noise mechanisms in the proposed circuit and their impact at the circulator's performance. Although the analysis focuses on the RX port, the results also apply to the ANT and TX ports, due to the symmetry of the circuit. Total noise at the RX port can be decomposed into three components: (i) incoming noise from the ANT and TX ports, (ii) noise added by the circuit itself, and (iii) noise resulting from random variations in the modulation signals, including amplitude and phase noise.

The TX and ANT noise are both transmitted to the RX port through the circulator's harmonic $S$-parameters. Notice that since the circuit is time-variant, output noise at the RX port at a particular frequency not only comes from the ANT or TX noise at the same frequency, but it also folds from the IM frequencies as shown in Fig. 13. Therefore, the RX noise due to the ANT and TX noise is given by

$$\overline{v_{RX,1}^2}(\omega) = \sum_{k=-N}^{N}\left[\begin{array}{l}\overline{v_{ANT}^2}(\omega + k\,\omega_m)\left|S_{21}(\omega, \omega + k\,\omega_m)\right|^2 \\ + \overline{v_{TX}^2}(\omega)\left|S_{31}(\omega, \omega + k\,\omega_m)\right|^2\end{array}\right], \quad (31)$$

where $N$ is the total number of the IM products and $\overline{v_{TX}^2}(\omega)$ and $\overline{v_{ANT}^2}(\omega)$ are the power spectral densities (PSDs) of the TX and ANT incoming noise, respectively. Notice that $\left|S_{31}(\omega, \omega)\right|^2 \square \left|S_{21}(\omega, \omega)\right|^2$ over the circulator's BW; therefore, TX's contribution at $k = 0$ in (31) can be neglected. Also, as mentioned earlier, only the second-order products at $f \pm f_m$ are excited under the linear small-signal assumption. Hence, (31) simplifies to

$$\overline{v_{RX,1}^2}(\omega) = \sum_{k=-1}^{1}\overline{v_{ANT}^2}(\omega + k\,\omega_m)\left|S_{21}(\omega, \omega + k\,\omega_m)\right|^2$$
$$+ \sum_{\substack{k=-1 \\ k\neq 0}}^{1}\overline{v_{TX}^2}(\omega + k\,\omega_m)\left|S_{31}(\omega, \omega + k\,\omega_m)\right|^2, \quad (32)$$

and the harmonic $S$-parameters in such a case are given by (18) -(25). One can argue that the circulator should not be penalized by the input noise folding, since there is no desired signal at the IM frequencies. A channel pre-selection bandpass filter, which is similar to image reject filters in heterodyne transceivers [31], can thus be added at the antenna port in order to knock down this out-of-band noise as shown in Fig. 13. Unlike DSB mixers, however, where the conversion gain seen by the desired signal and its image is the same, IM harmonic transfer functions in the circulator's case are, in fact, much smaller than insertion loss, i.e., $\left|S_{21,31}(\omega, \omega + k\,\omega_m)\right|_{k \neq 0} \square \left|S_{21}(\omega, \omega)\right|$, therefore, noise folding from IM frequencies is already very small compared to the in-band noise $N_{in}$ and can be neglected even without filtering.

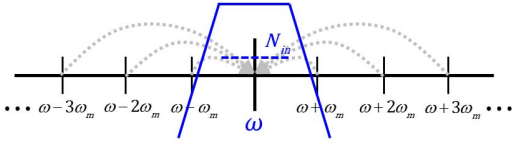

Fig. 13. Channel pre-selection filter to reject out-of-band input noise folding from IM frequencies.

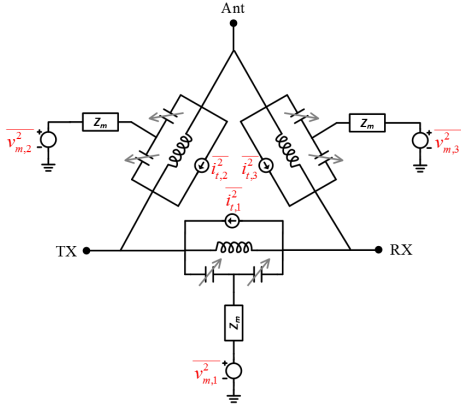

Fig. 14. Thermal noise sources generated by the circuit.

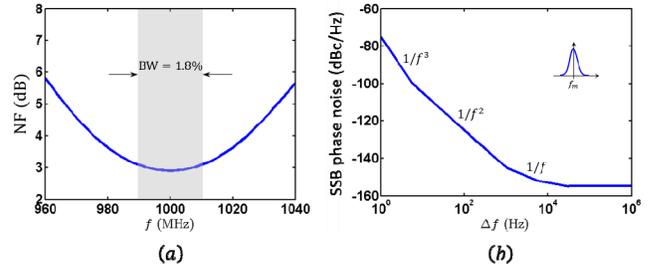

Fig. 15. (a) Simulated NF vs frequency. (b) Phase noise of the modulation sources.

As for the noise added by the circuit itself, it is attributed to the thermal noise of the biasing resistors, the inductors' finite quality factors, the varactors and the output impedance of the modulation sources. Fig. 14 shows these noise sources where noise of the biasing resistors $R_B$ is neglected compared to the input noise from the ports, since $R_B \square Z_0$. Also for simplicity, total noise of the $k$-th $LC$ tank, due to the varactors and the finite quality factor of $L$, is lumped into a single parallel current source $\overline{i_{t,k}^2}$. One can split each $\overline{i_{t,k}^2}$ into two fully correlated shunt current sources with opposite currents at the two terminals of the corresponding tank, therefore allowing the calculation of their contribution at the RX port using the harmonic $S$-parameters. Furthermore, incoming noise from the modulation/DC ports is injected into the common-cathode node of the corresponding tank through a voltage source $\overline{v_{m,k}^2}$ in series with a noiseless complex impedance $Z_m$ where $\overline{v_{m,k}^2}$ and $Z_m$ can be found using Thevenin-equivalence looking back into the modulation ports. One can interpret $\overline{v_{m,k}^2}$ as an amplitude variation of the modulation signal applied to the varactors, which is equivalent to an effective random capacitance variation $\delta C_n(t)$ in (5). Similarly, phase noise of

the modulation signals can be represented as a random phase $\theta_n(t)$, hence (5) is rewritten as follows:

$$C_n = C_0 + \left[\Delta C + \delta C_n(t)\right]\cos\left(\omega_m t + \varphi_n + \theta_n(t)\right). \quad (33)$$

Equation (33) shows that in a realistic scenario, the modulation signal is not a pure sinusoidal tone with frequency $f_m$, but a random signal with a finite bandwidth which increases the folded noise into the circulator's instantaneous BW and further degrades its NF. More importantly, noise in (33) at $f_m$ in particular is indistinguishable from the desired modulation signals at the same frequency, therefore, it would result in a random variation of the harmonic $S$-parameters, both magnitude and phase, hence the RF signal would incur undesired amplitude and phase modulation around the RF frequency. Clearly, this would lead to a fuzzy constellation in a practical communication system and increase the bit error rate, yet the significance of this effect depends on the communication scheme itself and is beyond the scope of this paper.

Analytical calculation of the exact total output noise at the RX port can be very tedious; however, simulation tools such as ADS harmonic balance significantly simplify the task. Also, since the presented circulator is a passive LPTV circuit, one should expect an overall noise figure (NF) close to the insertion loss [31] but slightly higher due to noise folding and modulation phase noise. Fig. 15(a) shows that the simulated NF is 2.9 dB at the center frequency, which is only 0.1 dB higher than IL, and less than 3.1 dB over the circulator's instantaneous BW. In obtaining these results, we assumed that the modulation sources are uncorrelated and have the phase noise characteristics shown in Fig. 15(b) in order to mimic the experimental setup we used in Section III.B.

## IV. MEASURED RESULTS

The circuit, as described in Section III (Fig. 2(a) for the schematic and Table II for the components), was fabricated on a PCB with a total form factor of 25mm×23mm and an area occupied by the core part (without the SMAs) of only 13mm×11mm. Fig. 16(a) and Fig. 16(b) show a photograph of the experimental setup and the fabricated prototype, respectively, while Table III provides a list of the used equipment.

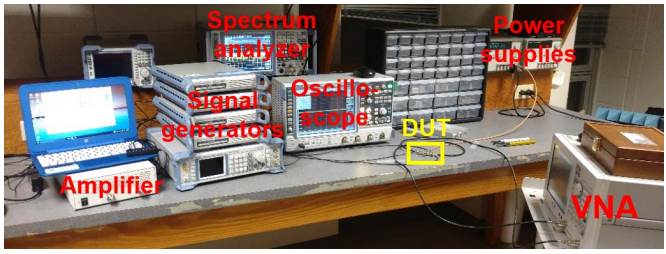

(a)

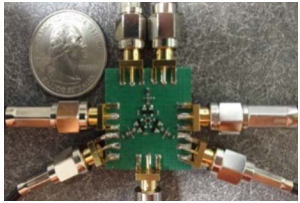

(b)

Fig. 16. Photograph of: (a) Experimental setup. (b) Fabricated prototype.

TABLE III
LIST OF USED EQUIPMENT

| Instrument | Model | Quantity |
|---|---|---|
| Power supply | Agilent E3631A | 1 |
| Vector network analyzer | Agilent E5071C | 1 |
| Spectrum analyzer | R&S FSVA40 | 1 |
| Oscilloscope | R&S RTO1044 | 1 |
| Signal generators | R&S SGS100A | 4 |
| | R&S SMB100A | 1 |
| Instrument amplifier | Minicircuits TVA-4W-422A+ | 1 |

### A. S-parameters and Output Spectrum

Fig. 17 shows a block diagram for the experimental setup used to measure the S-parameters. The modulation signals were generated using three phase-locked RF sources, which were controlled through a laptop. With the help of an oscilloscope, the sources were configured to generate three signals with the same amplitude and phase difference of 120 deg. Next, two of the RF ports were connected to the VNA and the third port was terminated with a 50 Ohm load. Ideally, a 4 port VNA would give the entire S-matrix with one measurement, while a 2 port VNA requires performing three different measurements in order to construct the $3 \times 3$ matrix. Finally, the DC port is connected to a power supply.

Fig. 18 shows the measured S-parameters, with and without modulation, for three different channels corresponding to different DC bias and modulation voltages. The modulation parameters (at the signal generators) in this case were chosen to get maximum isolation at the center frequency of each channel resulting in IL, RL and IX of 3.3 dB, −10.8, 55 dB, respectively, and 2.4% instantaneous BW (based on the definition of (27)). We also notice that matching is non-optimal, i.e., $S_{11}$ is not minimum at the circulator's center frequency. This is mainly due to the varactors' package uncertainities, which resulted in a self-resonance frequency (SRF) closer to 1 GHz than expected, thus leading to an asymmetric resonance of the unmodulated junction, as can be seen from $S_{11}$ in Fig. 18(a). It is worth mentioning that even with this asymmetry, the circulator is threefold rotationally symmetric from its ports, since the same S-parameters were measured between different ports, thus still preserving the insensitivity to port mismatches as compared to [24].

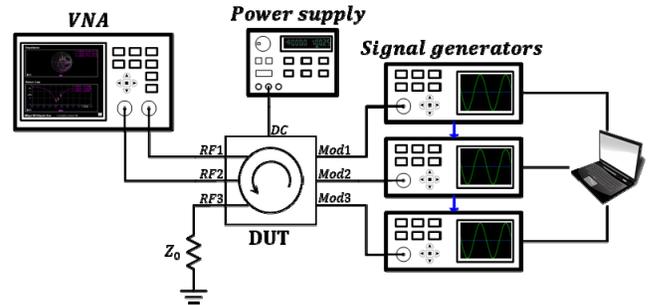

Fig. 17. Experimental setup of S-parameters measurements.

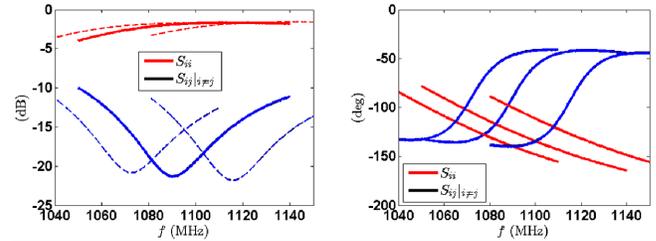

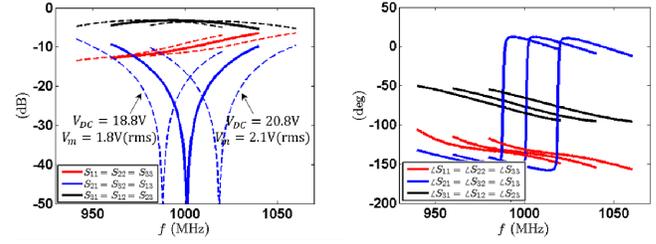

Fig. 18. Measured S-parameters: (a)-(b) Without modulation. (c)-(d) With modulation.

Fig. 19 shows the output spectrum at both the transmitted and isolated ports for a monochromatic input at $f_{rf} = 1$ GHz and $P_{in} = 0$ dBm. The output power $P_t$ is about −3.3 dBm which is indeed equal to $P_{in} - $IL as expected, while $P_{iso} = -48$ dBm is 7 dB larger than $P_{in} - $IX simply because the isolation null is not perfectly aligned with 1 GHz. We also see the second-order IM products at $f_{rf} - f_m$ and $f_{rf} + f_m$ which are −15 dBc and −11.3 dBc, respectively, in a fair agreement with the theoretical results. Furthermore, two additional tones exist at $5f_m$ and $6f_m$, which are basically higher-order harmonics of the modulation signal due to the varactors' non-linear $CV$ characteristics. It is therefore desirable not to have $f_{rf}/f_m$ as an integer number, in order to avoid any of these harmonics falling on top of the desired signal. Nevertheless, these harmonics are very small, e.g. the fifth harmonic is less than −40 dBc, and therefore can be neglected compared to the $f_{rf} \pm f_m$ IM products. Notice that these harmonics are independent of the RF signal and do not change when the input power is varied, therefore, they can be entirely cancelled out using simple DSP algorithms.

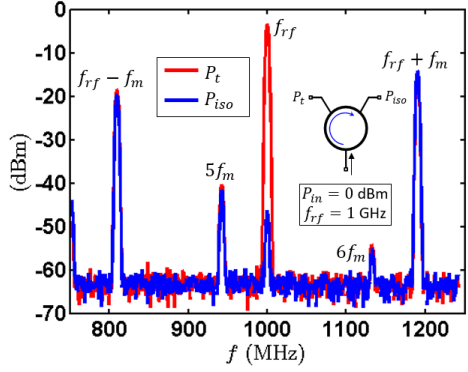

Fig. 19. Measured harmonic spectrum at the transmitted and isolated ports for a single tone input at $f_{rf} = 1$ GHz and $P_{in} = 0$ dBm.

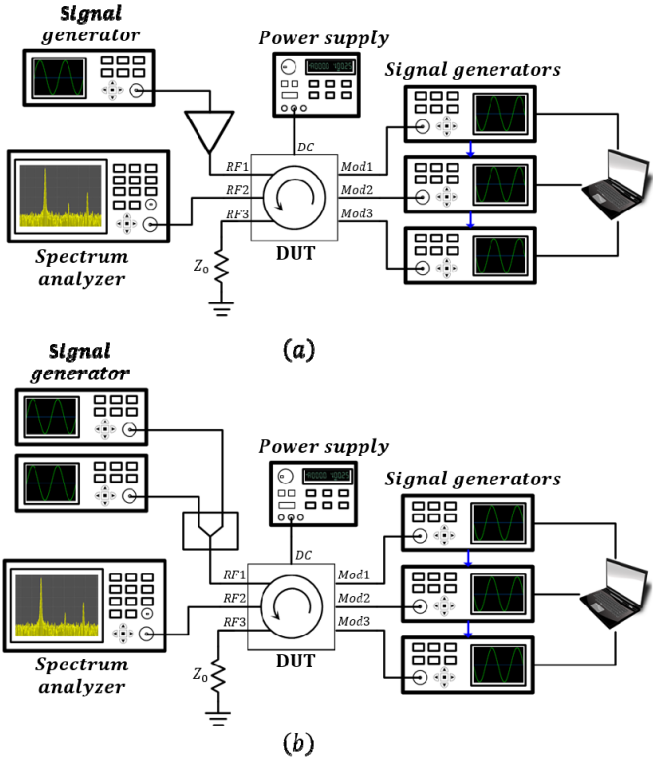

(a)

(b)

Fig. 20. Experimental setup of: (a) P1dB and (b) IIP3 measurements.

### B. Power Handling and Non-Linearities

In the measurement of P1dB, a monochromatic tone at $f_{rf} = 1$ GHz is generated via a signal generator and applied to port 1 of the circuit through an amplifier, as shown in Fig. 20(a). The amplifier was used in order to be able to tune the supplied power $P_{in}$ to the circuit at higher levels than the maximum output power of the generator. In this case, $P_{in}$ is restricted by the output compression point of the amplifier which is +34 dBm. The other two ports of the circuit are interchangeably connected to a spectrum analyzer and a 50 Ohm load in order to measure both transmission and isolation. The output powers $P_{TX}$ and $P_{IX}$ at the fundamental frequency of 1 GHz are measured using the spectrum analyzer and the measurement is repeated for different values of $P_{in}$ where the results are shown in Fig. 21. Both P1dB and $P_{max}$ (based on the definition of (30)) are found equal to 29 dBm which is a remarkable number for magnet-less circulators with such form factor.

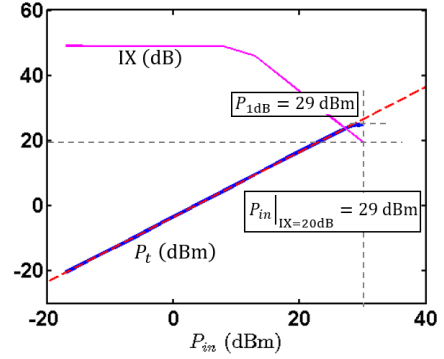

Fig. 21. Measured P1dB and maximum power handling.

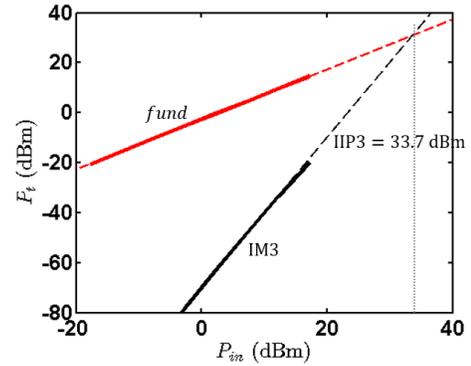

Fig. 22. Measured IIP3.

IIP3 is measured by combining two tones at $f_1 = 1000 + 0.5$ MHz and $f_2 = 1000 - 0.5$ MHz with the same power $P_{in}$ and feeding them to the circulator's input port, as shown in Fig. 20(b). Then the spectrum analyzer is used to measure the output power of the fundamental component at $f_1$ and the third-order IM product at $2f_1 - f_2$. Fig. 22 shows the results where IIP3 was found to be 33.7 dBm. To the best of the authors' knowledge, these are the largest reported P1dB and IIP3 of all magnet-less circulators proposed to date.

### C. Noise Figure

Fig. 23 shows a block diagram for the experimental setup used to measure the circulator's NF based on the Y factor method. A calibrated 6 dB excess noise ratio (ENR) source was connected to the circulator's input port and biased with a 28 V DC signal, the transmit port was connected to a spectrum analyzer through a high-pass filter (HPF) with a cut-off frequency larger than $f_m$ but much smaller than $f_{rf}$, and the isolated port was terminated to 50 Ohm. The HPF was added to reject any residual leakage from the modulation signals which, despite being small, may still require further attenuation to get below the noise floor in order not to overload the spectrum analyzer's RF port while measuring the NF. Notice that this filtering is only possible because $f_m$ is much smaller than $f_{rf}$ in the angular-momentum circulator, while if both frequencies were equal, e.g. as [24] requires, more complicated techniques must be utilized to reject the modulation leakage, which may, in turn, degrade the circulator's power handling.

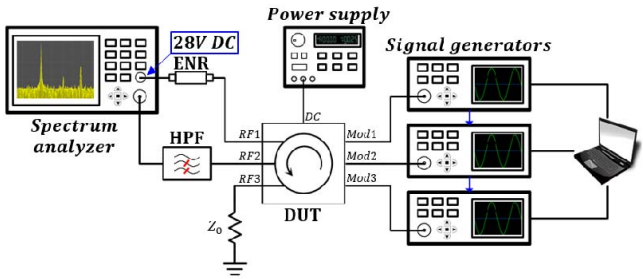

Fig. 23. Experimental setup of NF measurement.

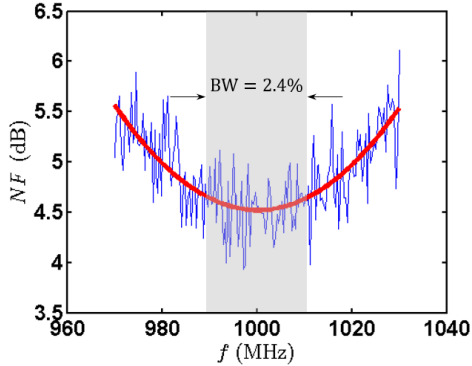

Fig. 24. Measured NF (in blue) and fitting curve (in red).

Fig. 24 shows the measured results when the circulator is configured to operate in the center channel of Fig. 18. In such a case, NF is 4.5 dB at 1 GHz and less than 4.7 dB over the circulator's instantaneous BW. These values were slightly larger than predicted in simulations mainly because the measured insertion loss was 0.5 dB higher. It is also worth mentioning that the phase noise contribution to the NF can be reduced by using one source and generating the three modulation signals on-board using phase shifters such that they become perfectly correlated, hence the random phase variations of the generator's output signal equally appear in the three modulation signals.

## V. Conclusion

We presented here the concept, design principles and experimental verification of magnet-less circulators based on spatio-temporally modulated first-order bandstop filters connected in a delta topology. We developed a rigorous analytical model, from which we extracted the optimal modulation parameters of the circuit in order to achieve given specifications on insertion loss, return loss, isolation and bandwidth. Based on this model, we designed a PCB prototype and measured its performance including scattering parameters, harmonic response, power handling, and noise figure as summarized together with the theoretical and simulated results in Table IV in comparison with previous works. These results show that the proposed circuit is very promising for the realization of tunable and IC compatible magnet-less circulators with small form-factor, low loss, large isolation, high power handling, and low NF, as required in practical full-duplex systems, and has definite advantages compared to other recently proposed approaches to magnet-free circulators.

TABLE IV
SUMMARY OF RESULTS AND COMPARISON TO PREVIOUS WORK

| Metric | This work | | | [22] Meas. | [23] Meas. |
|---|---|---|---|---|---|
| | Theo. | Sim. | Meas. | | |
| RF frequency (MHz) | 1000 | 1000 | 1000 | 170 | 130 |
| DC bias (Volt) | N.A. | 21.6 | 19.6 | 2 | 1.7 |
| Mod. frequency (MHz) | 190 | 190 | 190 | 15 | 40 |
| Mod. source voltage (rms) | N.A. | 2.5 | 3.8 | 0.42 | 0.9 |
| DC power consumption (mW) | 0 | 0 | 0 | 0 | 0 |
| Isolation[1] (dB) | 56 | 55 | 55 | 40 | 50 |
| Insertion loss[1] (dB) | 2.9 | 2.8 | 3.3 | 25 | 9 |
| Return loss[1] (dB) | −10.8 | −11.3 | −10.8 | N.A. | 4 |
| Instantaneous BW[2] (%) | 2.7 | 1.8 | 2.4 | N.A. | N.A. |
| Tunability BW (%) | 10 | 10 | 6 | 30 (*) | N.A. |
| P1dB (dBm) | N.A. | 28 | 29 | N.A. | N.A. (**) |
| IIP3 (dBm) | N.A. | 33.8 | 33.7 | N.A. | N.A. (***) |
| Pmax[3] (dBm) | N.A. | 26.7 | 29 | N.A. | N.A. |
| Max. IM product (dBc) | −10 | −10 | −11.3 | N.A. | N.A. |
| Modulation leakage (dBc) | N.A. | −60 | −60 | N.A. | N.A. |
| NF[1] (dB) | N.A. | 2.9 | 4.5 | N.A. | N.A. |
| Size (mm²) | N.A. | 13 × 11 | 13 × 11 | 25 × 25 | 22 × 17 |

[1] At center frequency
[2] Defined in (27)
[3] Defined in (30)
*Applies only for IX
**P1dB(sim) = +3.2 dBm
***IIP3(sim) = +5.7 dBm

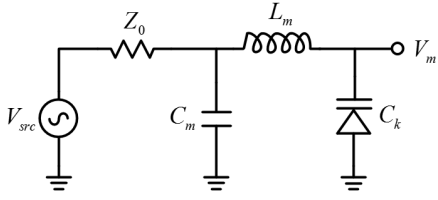

Fig. 25. $L$-section matching network of the $k$-th modulation branch.

Several of these metrics, including BW, insertion loss and unwanted intermodulation products, can be further improved using higher-order modulated filters or a differential architecture, which will be the subject of future investigations.

## APPENDICES

### A. Modulation Network and DC Biasing

As mentioned in Section II.B, the modulation signals *see* a virtual ground at the RF ports because of symmetry. To prove this fact, we consider $f_m \square f_{rf}$, as is the case in this paper. In such a case, the impedance of the RF inductors $L$ at $f_m$ is small enough that they can be replaced with a short circuit. Therefore, each pair of common cathode varactors in Fig. 2(a) appear in parallel and can be effectively replaced with a single varactor as shown in Fig. 2(c). Similarly, the internal impedance $Z_0$ of the three RF ports also appear in parallel and are lumped into a single resistance $R_{eq} = Z_0 / 3$. Because of this resistance, the modulation voltage that is applied across the varactors and enters in (4) for the effective capacitance seen by the RF signal is generally not equal to the modulation voltages $v_{mk}$ after the modulation filters, but $v_{mk} - v_p$, where $v_p$ is the voltage at the modulation frequency across $R_{eq}$. However, due to the 120 deg symmetry of the modulation signals at the three varactors, the current flowing through $R_{eq}$ is zero, making the node P a virtual ground for the modulation signals and the modulation signals across the varactors equal to $v_{mk}$.

The virtual ground feature simplifies the design of the modulation network since the circuit in Fig. 2(c) can now be split into three identical and independent circuits as shown in Fig. 25 for the $k$th branch. Such a circuit can be designed to amplify the voltage at the varactor compared to the source voltage, thus relaxing the requirements on the modulation signal generators $V_{src}$. By using simple circuit analysis, the voltage gain $G_m$ can be found as

$$G_m = \frac{V_m}{V_{src}} = \left| 1 + R_{src}\omega_m \left[ \frac{1 - \omega_m^2 L_m C_{eq}}{1 - \omega_m^2 L_m C_k} \right] \right|^{-1}, \quad (34)$$

where $R_{src}$ is the finite output impedance of the modulation sources, $L_m$ and $C_m$ are the matching network's series inductance and shunt capacitance, respectively, and $C_{eq} = \frac{C_m C_k}{C_m + C_k}$. The condition to get the maximum gain is given by

$$\left( \frac{1 - \omega_m^2 L_m C_{eq}}{1 - \omega_m^2 L_m C_k} \right) = -\frac{1}{R_{src}\omega_m}. \quad (35)$$

Also, in order to prohibit the RF signal from leaking into the modulation ports, the following condition must be satisfied:

$$L_m > 20L, \quad (36)$$

such that $\omega_{rf} L_m$ is large enough to be considered an open circuit for the RF signal. Equations (35) and (36) can now be used to find the values of $L_m$ and $C_m$ for optimal operation of the circulator. Finally, the DC bias can be combined with the modulation signals through a bias tee or a sufficiently large resistance, as $R_B$ in Fig. 2(a), since there is no DC current flow in the circuit. Furthermore, for proper biasing of the varactors, the RF ports are also DC grounded again through a bias tee or a large resistance.

### B. Detailed Small-Signal Analysis

Here, we present a detailed analysis for the small-signal model of Fig. 2(b) and derive the equations given in section II.B. Applying Kirchhoff's laws to the $n$th tank in Fig. 2(b), we get

$$v_n = Li'_{L,n}, \quad (37)$$

$$i_{L,n} = i_n - \frac{v_n}{R} - C_n v'_n, \quad (38)$$

where $' = \frac{d}{dt}$, $i_{L,n}$ is the current in the inductor of the $n$-th tank, $i_n$ is the total current in the $n$-th branch (see Fig. 2(b)) and $n = 1,2,3$. Taking the derivative of (38) and substituting it into (37), we get

$$i'_n = C_n v''_n + \left( \frac{1}{R} + C'_n \right) v'_n + \frac{1}{L} v_n. \quad (39)$$

Substituting (5) into (39), we get

$$i'_n = \left[ C_0 + \Delta C \cos\left(\omega_m t + \varphi_n\right) \right] v''_n + \left[ \frac{1}{R} - \Delta C \omega_m \sin\left(\omega_m t + \varphi_n\right) \right] v'_n + \frac{1}{L} v_n. \quad (40)$$

The voltages $v_n$ can also be related to the source voltages $v_{s1,2,3}$ at the RF ports as follows

$$v_1 = \left[ v_{s1} - Z_0\left(i_1 - i_3\right) \right] - \left[ v_{s2} - Z_0\left(i_2 - i_1\right) \right], \quad (41)$$

$$v_2 = \left[ v_{s2} - Z_0\left(i_2 - i_1\right) \right] - \left[ v_{s3} - Z_0\left(i_3 - i_2\right) \right], \quad (42)$$

$$v_3 = \left[ v_{s3} - Z_0\left(i_3 - i_2\right) \right] - \left[ v_{s1} - Z_0\left(i_1 - i_3\right) \right]. \quad (43)$$

Substituting (41)-(43) into (40) and using $v_1 + v_2 + v_3 = 0$ from Kirchhoff's voltage law, we get three differential equations for $v_n$, which were compactly written in the matrix form (6) where

$$\overline{\overline{A}} = Z_0\left(3C_0\overline{\overline{U}} + \Delta C\overline{\overline{C_c}}\right), \quad \overline{\overline{B}} = \overline{\overline{U}} + Z_0\left(\frac{1}{R}\overline{\overline{H}} - \Delta C\,\omega_m\overline{\overline{C_s}}\right), \quad \overline{\overline{U}} \text{ is}$$

the unitary matrix, and

$$\overline{\overline{G}} = \begin{bmatrix} +1 & -1 & 0 \\ 0 & +1 & -1 \\ -1 & 0 & +1 \end{bmatrix}, \tag{44}$$

$$\overline{\overline{H}} = \begin{bmatrix} +2 & -1 & -1 \\ -1 & +2 & -1 \\ -1 & -1 & +2 \end{bmatrix}, \tag{45}$$

$$\overline{\overline{C_c}} = \begin{bmatrix} 2\cos(\omega_m t) & -\cos(\omega_m t + \alpha) & -\cos(\omega_m t + 2\alpha) \\ -\cos(\omega_m t) & 2\cos(\omega_m t + \alpha) & -\cos(\omega_m t + 2\alpha) \\ -\cos(\omega_m t) & -\cos(\omega_m t + \alpha) & 2\cos(\omega_m t + 2\alpha) \end{bmatrix}, \tag{46}$$

$$\overline{\overline{C_s}} = -\frac{1}{\omega_m}\overline{\overline{C_c}}' =$$
$$\begin{bmatrix} 2\sin(\omega_m t) & -\sin(\omega_m t + \alpha) & -\sin(\omega_m t + 2\alpha) \\ -\sin(\omega_m t) & 2\sin(\omega_m t + \alpha) & -\sin(\omega_m t + 2\alpha) \\ -\sin(\omega_m t) & -\sin(\omega_m t + \alpha) & 2\sin(\omega_m t + 2\alpha) \end{bmatrix}. \tag{47}$$

Applying the matrix transformation (7),(8) to (6) yields

$$\tilde{\overline{\overline{A}}}\vec{v}'' + \tilde{\overline{\overline{B}}}\vec{v}' + \frac{3Z_0}{L}\vec{v} = \tilde{\overline{\overline{G}}}\vec{v}_s', \tag{48}$$

where

$$\tilde{\overline{\overline{A}}} = \overline{\overline{T}}^{-1}\overline{\overline{A}}\overline{\overline{T}} =$$
$$3Z_0\begin{bmatrix} C_0 & 0 & 0 \\ \dfrac{\Delta C}{2}e^{j\omega_m t} & C_0 & \dfrac{\Delta C}{2}e^{-j\omega_m t} \\ \dfrac{\Delta C}{2}e^{-j\omega_m t} & \dfrac{\Delta C}{2}e^{j\omega_m t} & C_0 \end{bmatrix}, \tag{49}$$

$$\tilde{\overline{\overline{B}}} = \overline{\overline{T}}^{-1}\overline{\overline{B}}\overline{\overline{T}} =$$
$$\begin{bmatrix} 1 & 0 & 0 \\ \dfrac{j3}{2}Z_0\Delta C\omega_m e^{j\omega_m t} & 1 + \dfrac{3Z_0}{R} & \dfrac{-j3}{2}Z_0\Delta C\omega_m e^{-j\omega_m t} \\ \dfrac{-j3}{2}Z_0\Delta C\omega_m e^{-j\omega_m t} & \dfrac{j3}{2}Z_0\Delta C\omega_m e^{j\omega_m t} & 1 + \dfrac{3Z_0}{R} \end{bmatrix}, \tag{50}$$

$$\tilde{\overline{\overline{G}}} = \overline{\overline{T}}^{-1}\overline{\overline{G}} = \begin{bmatrix} 0 & 0 & 0 \\ \dfrac{1}{6}(3 - j\sqrt{3}) & -\dfrac{1}{6}(3 + j\sqrt{3}) & \dfrac{j}{\sqrt{3}} \\ \dfrac{1}{6}(3 + j\sqrt{3}) & -\dfrac{1}{6}(3 - j\sqrt{3}) & -\dfrac{j}{\sqrt{3}} \end{bmatrix}. \tag{51}$$

For $\vec{v}_s = \{1, 0, 0\}$, (48) can be decomposed into the three equations (9)-(11). The solution of (9) is $v_{cm} = 0$. On the other

hand, in order to find the solutions of (10), (11), we apply the Fourier transform, yielding

$$-3Z_0C_0\omega^2V_+(\omega) - \frac{3}{2}Z_0\Delta C(\omega + \omega_m)^2V_-(\omega + \omega_m)$$
$$+j\left(1 + \frac{3Z_0}{R}\right)\omega V_+(\omega) + \frac{3}{2}Z_0\Delta C\omega_m(\omega + \omega_m)V_-(\omega + \omega_m) \tag{52}$$
$$+\frac{3Z_0}{L}V_+(\omega) = \frac{j}{6}\left(3 - j\sqrt{3}\right)\omega V_{s1}(\omega),$$

$$-3Z_0C_0\omega^2V_-(\omega) - \frac{3}{2}Z_0\Delta C(\omega - \omega_m)^2V_+(\omega - \omega_m)$$
$$+j\left(1 + \frac{3Z_0}{R}\right)\omega V_-(\omega) - \frac{3}{2}Z_0\Delta C\omega_m(\omega - \omega_m)V_+(\omega - \omega_m) \tag{53}$$
$$+\frac{3Z_0}{L}V_-(\omega) = \frac{j}{6}(3 + j\sqrt{3})\omega V_{s1}(\omega),$$

where $V_+(\omega)$ and $V_-(\omega)$ are the Fourier transforms of $v_+(t)$ and $v_-(t)$, respectively. Solving (52) and (53) for $V_+(\omega)$ and $V_-(\omega)$ leads to (12)-(17).

The current flowing in each branch can also be found by transforming (39) to frequency domain, resulting in

$$I_n(\omega) = \left(\frac{1}{R} + j\omega C_0 + \frac{1}{j\omega L}\right)V_n(\omega)$$
$$+\frac{j}{2}\Delta C\begin{bmatrix} e^{j(n-1)\alpha}(\omega - \omega_m)V_n(\omega - \omega_m) \\ +e^{-j(n-1)\alpha}(\omega + \omega_m)V_n(\omega + \omega_m) \end{bmatrix}, \tag{54}$$

$$I_n(\omega + \omega_m) = \frac{j}{2}\Delta Ce^{j(n-1)\alpha}\omega V_n(\omega) +$$
$$\left(\frac{1}{R} + j(\omega + \omega_m)C_0 + \frac{1}{j(\omega + \omega_m)L}\right)V_n(\omega + \omega_m), \tag{55}$$

$$I_n(\omega - \omega_m) = \frac{j}{2}\Delta Ce^{-j(n-1)\alpha}\omega V_n(\omega)$$
$$+\left(\frac{1}{R} + j(\omega - \omega_m)C_0 + \frac{1}{j(\omega - \omega_m)L}\right)V_n(\omega - \omega_m). \tag{56}$$

Finally, the harmonic $S$-parameters are calculated using (24) and (25).